\documentclass[12pt]{article}

\usepackage{epsfig}
\usepackage{scicite}
\usepackage{sidecap}

\usepackage{times}

\newenvironment{sciabstract}{%
\begin{quote} \bf}
{\end{quote}}

\newcounter{lastnote}

\usepackage[T1]{fontenc}
\usepackage{mathptmx}
\usepackage{hyperref}

\hypersetup{
   colorlinks=true,
   citecolor=blue,
   linkcolor=blue,
   urlcolor=black
   }
   
\def\beq{\begin{equation}}
\def\eeq{\end{equation}}
\usepackage{amsmath,amssymb,amsbsy}

\def\brunt{Brunt-Väisälä }

\def\rp{R}
\def\Mp{M}

\def\mearth{M_\oplus}

\def\cm2s{{\rm cm}^2\,{\rm s}^{-1}}
\def\ga{\,\hbox{\hbox{$ > $}\kern -0.8em \lower 1.0ex\hbox{$\sim$}}\,}
\def\la{\,\hbox{\hbox{$ < $}\kern -0.8em \lower 1.0ex\hbox{$\sim$}}\,}

\def\d{\mathrm{d}}

\def\op{\omega}
\def\hp{H_P}
\def\hl{l}
\def\delt{\nabla_T}
\def\delad{\nabla_\mathrm{ad}}

\def\deld{\nabla_\mathrm{d}}
\def\delmean{\langle\nabla_T\rangle}

\def\deltat{\delta_T}

\def \kapt{\kappa_T}

\def \tKH{\tau_\mathrm{KH}}

\def\angmom{J}
\def\anginertia{I}

\def\ttb{T_\mathrm{10}}
\def\te{T_{\rm eff}}
\def\tint{T_{\rm int}}
\def\teq{T_{\rm eq}}
\def\tmid{T_\mathrm{mid}}

\def\fint{F_\mathrm{int}}

\def\egrav{E_\mathrm{g}}
\def\eth{E_\mathrm{th}}
\def\erot{E_\mathrm{rot}}
\def\etot{E_\mathrm{tot}}
\def\uint{\tilde{u}}

\def\Lint{L_\mathrm{int}}
\def\Lsun{L_\odot}

\def\ssb{\sigma_{SB}}

\def\Nt{N_T}

\newcommand{\itref}[1]{$^{\ref{#1}}$}

\newcommand{\pd}[2]{\frac{\partial \!\! \ #1}{\partial \!\! \ #2}}

\newcommand{\pdc}[3]{\left. \frac{\partial \!\! \ #1}{\partial \!\! \ #2}\right|_{#3}}

\newcommand{\balign}[1]{
\begin{align}
#1
\end{align}}

\newcommand{\eq}[1]{Eq.\,(\ref{#1})}

\newcommand{\fig}[1]{Fig.\,\ref{#1}}

\newcommand{\tab}[1]{Table\,\ref{#1}}

\title{ Layered convection as the origin of \\Saturn's luminosity anomaly}

\author{J\'er\'emy Leconte$^{1,2}$ \& Gilles Chabrier$^{2,3}$\\
\\
\normalsize{$^{1}$Laboratoire de M\'et\'eorologie Dynamique, Institut Pierre Simon Laplace, Paris, France,}\\
\normalsize{$^{2}$Ecole Normale Sup\'erieure de Lyon, CRAL (UMR CNRS 5574), 69364 Lyon, France,}\\
\normalsize{$^{3}$School of Physics, University of Exeter, Exeter, UK}\\
\\
\normalsize{E-mail:  jeremy.leconte@lmd.jussieu.fr, chabrier@ens-lyon.fr}
}

\date{}
\newpage

\begin{document} 


\baselineskip14pt


\maketitle 

\begin{sciabstract}

As they keep cooling and contracting, Solar System giant planets radiate more energy than they receive from the Sun. 
Applying the first and second principles of thermodynamics, one can determine their cooling rate, luminosity, and temperature at a given age. Measurements of Saturn's infrared intrinsic luminosity, however, reveal that this planet is significantly brighter than predicted for its age\itref{PGM77}$^,$\itref{FIN11}. This excess luminosity is usually attributed to the immiscibility of helium in the hydrogen-rich envelope, leading to "rains" of helium-rich droplets\itref{Sal73}$^-$\itref{MSC09}. Existing evolution calculations, however, suggest that the energy released by this sedimentation process may not be sufficient to resolve the puzzle\itref{FH03}.
Here, we demonstrate using planetary evolution models that the presence of layered convection in Saturn's interior, generated, like in some parts of Earth oceans, by the presence of a compositional gradient, significantly reduces its cooling. It can explain the planet's present luminosity for a wide range of configurations without invoking any additional source of energy.
This suggests a revision of the conventional homogeneous adiabatic interior paradigm for giant planets, and questions our ability to assess their heavy element content. This reinforces the possibility for layered convection to help explaining the anomalously large observed radii of extrasolar giant planets.

 \end{sciabstract}

Many arguments suggest the existence of compositional gradients in giant planet interiors, as a consequence of either their formation process or their cooling history\itref{Ste85}$^,$\itref{PHS91}$^,$\itref{CB07}$^,$\itref{LC12}.
However, the effect of this gradient on the thermal evolution of the planet is usually neglected for sake of simplicity.
When a vertical gradient of heavy elements is present in a fluid, the resulting mean molecular weight gradient (decreasing upward) can prevent large scale convection  to develop by counteracting the destabilizing effect of the temperature gradient.
The complex interaction between advection and diffusion of heat and atomic concentration can trigger a hydrodynamical instability, called
the double diffusive instability\itref{Ste60}, leading to a regime of double diffusive convection (also called semi convection). 
This process significantly affects heat and element transport, as observed in Earth oceans. The instability leads to either a state of homogeneous double diffusive convection where diffusive transport is only modestly enhanced by small scale turbulence, or to a state of "layered convection", with numerous, small convective layers separated by thin, diffusive interfaces corresponding to discontinuities in the composition of the fluid, in which transport is enhanced more significantly\itref{Rad05}$^,$\itref{RGT11}. 

Numerical calculations show that, for relevant thermal and atomic diffusivities, both scenarios are possible, depending on the ratio of the compositional to superadiabatic temperature gradients\itref{MGS12}. In planets, however, this latter quantity is not a free parameter but is imposed by the (measured) energy flux to be transported in the planet. It can be shown that this flux is too high to be transported by the diffusive regime of the double diffusive instability (said differently, the thermal gradient needed to transport the internal flux over the whole planet would be too high to be stabilized by the heavy element gradient\itref{LC12}). 
Under planetary conditions, the system will thus most likely settle in the regime of layered convection.


To study the impact of layered convection on giant planet thermal structure, we have developed an analytical model of the heat transport in such a medium\itref{LC12}. Extending the usual mixing length formalism\itref{HK94} to layered convection, we can calculate the internal temperature gradient as a function of the local properties of the fluid and of a unique free parameter, namely the characteristic size of the convective layers, $\hl$, or equivalently the corresponding dimensionless mixing length parameter, $\alpha=\hl/\hp$, where $\hp$ is the pressure scale height in the fluid (see Methods).

\begin{figure}[htbp] 
 \centering
 \resizebox{.6\hsize}{!}{\includegraphics{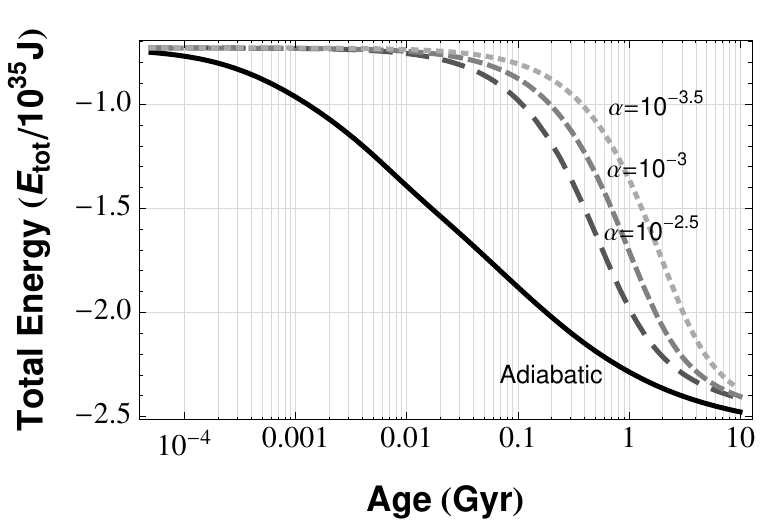}}
 \caption{
Evolution of the total internal energy ($\etot\equiv\egrav + \eth + \erot $) with time for various Saturn models. The solid curve corresponds to the reference adiabatic model. The long dashed, dashed and dotted curves are models with layered convection with $\alpha= 10^{-2.5}$, $10^{-3}$ and $10^{-3.5}$, respectively.  Because of the reduced intrinsic luminosity caused by layered convection, these models cool more slowly and keep the memory of their initial energetic state much longer.}
 \label{fig:EtotSat}
\end{figure}

A first confirmation of the viability of this scenario for giant planet interiors has been given by the fact that structure models with layered convection matching all the observational \textit{mechanical} constraints of Jupiter and Saturn (radius, gravitational moments), as well as the atmospheric helium and mean heavy element abundances, have been obtained for a rather wide range of mixing length parameters, namely $\alpha \in [10^{-2}-4\times10^{-6}]$ for Saturn and $\alpha \in [10^{-2}-3\times10^{-5}]$ for Jupiter\itref{LC12}. These constraints suggest that layered convection is favored over homogeneous double diffusive convection (Supplementary Information). However, the question remains whether layered convection can also explain the present luminosity of the giant planets.


In order to answer this question, we have computed the thermal evolution of planets with layered convection. Because departure from adiabaticity can be significant in these interiors, usual isentropic evolution calculations\itref{FIN11} cannot be used. Instead, the evolution is computed by integrating the intrinsic luminosity, $\Lint=-\d\, \etot/ \d\, t$, where the total energy $\etot=\egrav+\eth+\erot$ includes the gravitational, thermal and rotational energies (see Methods). The size of the convective/diffusive layers is assumed to have reached an equilibrium value\itref{LC12}$^,$\itref{Rad05} and is thus kept constant. For all the models (i.e. the reference, homogeneous/adiabatic case and the semi convective ones, for any given $\alpha$), the amount of heavy elements is kept constant and equal to the one that fulfills the gravitational moment constraints\itref{LC12} (Supplementary Information).

\begin{figure}[htbp] 
 \centering
 \resizebox{.6\hsize}{!}{\includegraphics{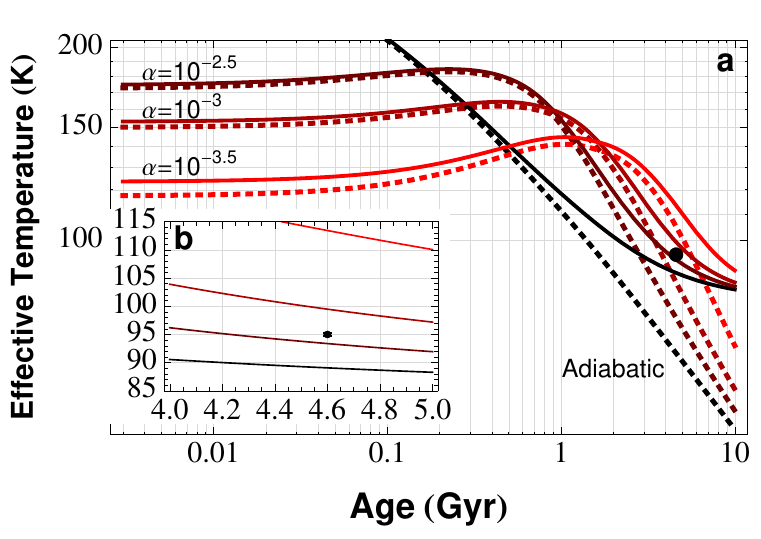}}
 \caption{
Cooling sequences of Saturn models with layered convection. a) Effective ($\te$; solid curves) and intrinsic ($\tint$; dashed curves) temperature (see Methods) evolution in time for the adiabatic reference model (black) and three models with layered convection ($\alpha= 10^{-2.5}$, $10^{-3}$ and $10^{-3.5}$ from dark to light red). b) Zoom on present era. Dots show the observed effective temperature. At early ages the effective temperature of models with layered convection is lower due to inefficient convection. After a few hundred million years, these models become brighter due to the release of the excess of energy stored from the initial state. }
 \label{fig:TeffSaturn}
\end{figure}

\begin{figure}[htbp] 
 \centering
 \resizebox{.6\hsize}{!}{\includegraphics{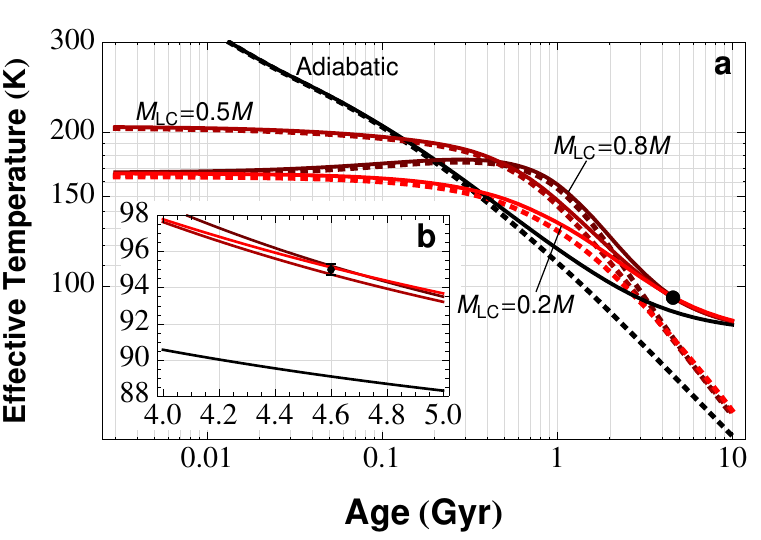}}
 \caption{
Impact of the size of the layered convection zone on Saturn's cooling sequence. a) Effective ($\te$; solid curves) and intrinsic ($\tint$; dashed curves) temperature evolution for the adiabatic model (black curve) and three scenarios with layered convection. b) Zoom on the present era. From darker to lighter red curves we have, i) our baseline scenario with layered convection present throughout the gaseous envelope of the planet ($M_\mathrm{LC}=0.8\Mp$; $\alpha= 2\times10^{-3}$), ii) a scenario where only the inner 50\% of the envelope exhibits layered convection ($M_\mathrm{LC}\approx0.5\,\Mp$; $\alpha=2\times10^{-4}$), iii) a case with a layered convection shell between 0.5 and 0.7$\,\Mp$ ($M_\mathrm{LC}\approx0.2\,\Mp$; $\alpha=10^{-4}$). 
}
 \label{fig:TeffSaturn_Msc}
\end{figure}

As illustrated in \fig{fig:EtotSat}, our results show that starting from the same total internal energy, the time required to release a given amount of this energy is significantly longer in the layered case than in the homogeneous adiabatic one.
This does not imply, however, that the intrinsic luminosity of the object will necessarily be larger at any given time. On the contrary, at early ages, objects with a layered convection zone are far less luminous because of the reduced heat flux release.
However, after some time, the decrease in luminosity imposed by layered convection is overpowered by the increase of internal energy to be released and planets with layered convection eventually become more luminous than the ones with adiabatic interiors (\fig{fig:TeffSaturn}). In Saturn's case, this "crossing time" occurs before Saturn's present age so that
layered convection yields a larger luminosity at 4.6 Gyr. As shown in \fig{fig:TeffSaturn} and \fig{fig:TeffSaturn_Msc}, if the size of the convective layers is about 2-3$\times 10^{-3}$ pressure scale heights,  this process properly leads to the currently observed effective temperature, radius, and gravitational moments of the planet without any additional energy source such as helium rains. The radius evolution of these models is shown in Supplementary \fig{fig:ReqSat}.
For this range of layer sizes, the present interior temperature of Saturn is only modestly affected compared to the adiabatic case (see Supplementary \fig{fig:StructureSat}). The temperature during the early evolution, however, is much higher.

\fig{fig:TeffSaturn} suggests that Saturn models with convective/diffusive layers smaller than 2-3$\times 10^{-3}$ pressure scale heights will be too bright at the age of the Solar System.
This is not necessarily true. If layered convection develops only within a restricted fraction of the planet, models with a lower convective efficiency (lower $\alpha$) in the layered zone and with an efficient convection everywhere else can also reproduce Saturn's proper cooling timescale. 
Our results are valid for very different sizes of the layered convection zone. Without covering the whole parameter space, we illustrate this by showing an evolution track for Saturn where layered convection is restricted to the inner 50\% in mass of the planet above the core, with $\alpha=2\times 10^{-4}$, and an other track where layered convection occurs in a shell between 50\% and 70\% of the planet's mass
(\fig{fig:TeffSaturn_Msc}). This latter case shows that the layered convection region does not need to extend down to the core and could be present around the molecular/metallic transition region or near an immiscibility region in the gaseous envelope.


We have also tested the sensitivity of the model to the initial conditions. 
Indeed, scenarios of giant planet formation by core accretion do not provide so far a clear prescription for these initial conditions. 
Large uncertainties in various parameters (gas-to-dust ratio, opacities,...) and the lack of a proper treatment of the radiative shock during the gas runaway accretion phase, for instance, hamper a proper determination of the planet's initial energy content\itref{PHB96}$^,$\itref{MFH07}. 
Whereas these (unknown) initial conditions become inconsequential after about a few tens of million years for an adiabatic planet of Saturn's mass\itref{MFH07}, it takes significantly longer (up to a few billion years) for a planet where convection is inefficient, because of the lower luminosity and thus the longer thermal, 
so-called Kelvin-Helmholtz timescale ($\tau_{KH}\propto 1/\Lint$).
Initial conditions thus have a stronger impact on the cooling timescale of planets with layered convection.

\begin{figure}[htbp] 
 \centering
 \resizebox{.6\hsize}{!}{\includegraphics{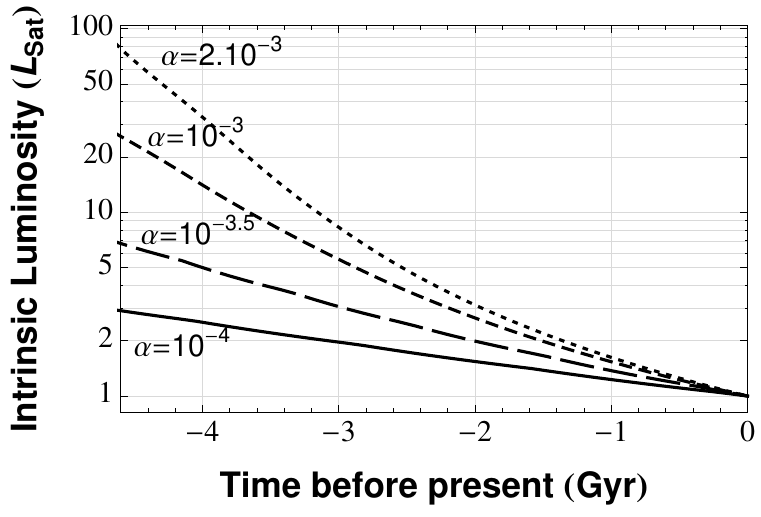}}
 \caption{Past luminosity of Saturn's model with layered convection. Evolutionary tracks of the intrinsic luminosity (in units of the presently measured luminosity) of Saturn models integrated backward in time from presently observed conditions with different mixing length parameters ($\alpha$). All the models with $\alpha>2\times10^{-3}$ have too short cooling timescales and cannot be integrated backward for 4.6\,Gyr. Even for quite low initial energy contents, i.e. a low initial intrinsic luminosity, models with a sufficiently inefficient convection can reproduce the currently observed luminosity of Saturn. }
 \label{fig:Einit_alpha}
\end{figure}

As shown in \fig{fig:Einit_alpha}, however, these different initial conditions do not qualitatively affect our results, as there is always a possible trade-off between the initial energy content of the nascent planet (at the end of the runaway accretion shock) and the thickness or number of convective-diffusive layers. As seen in the figure, evolution of models with layered convection can adequately reproduce Saturn's presently observed luminosity with very low initial energy content provided sufficiently small convective/diffusive layers.
Given the current uncertainties of formation models, 
thermal evolution of Saturn with layered, inefficient convection can thus meet the observational constraints for a wide variety of initial conditions.
 Interestingly enough, the same layered convection mechanism, which inhibits the heat flow, also provides a plausible explanation to the observed low luminosity of Uranus\itref{PHS91}. In this case, however, the "crossing time" discussed above should be larger than the age of the Solar System.

%

Because the present luminosity of Jupiter is smaller than the one that is predicted by homogeneous, adiabatic evolution models\itref{FIN11} (see Supplementary \fig{fig:TeffJupSat}), 
observations do not seem to support the presence of layered convection over the whole planet at the present epoch. 
This does not preclude, however, the possibility for this process either to be present in part of the planet or to have occurred in the past, but the planet's gaseous envelope to have been homogenized early enough, 
due to the more vigorous convective flux. The flux emitted by Jupiter during its cooling history is large enough to redistribute a significant fraction of a massive initial core 
(of up to $15-20\,\mearth$) within less than 4.6\,Gyr\itref{GSS}. 
Layered convection in this planet could then have stopped after the erosion of the whole core, or at least of its soluble components\itref{WM12}$^,$\itref{WM12b}. 
This would be consistent with both i) the large initial core mass needed to form the giant planet on the right timescale\itref{Ste82} and 
ii) Jupiter's present smaller core\itref{GSS}$^,$\itref{SG04}$^,$\itref{LC12}. Jupiter and Saturn may thus have experienced quite different cooling histories and end up with dissimilar interiors.

Layered convection does not preclude either the possibility of a phase separation between hydrogen and either volatiles or helium in Saturn's interior. In fact, a phase separation favors the occurrence of layered convection\itref{SS77a}.
Indeed, the strong compositional gradients (discontinuities) due to the presence of an immiscibility region provide very favorable conditions for the development of the double diffusive instability and thus of layered convection. Our present treatment of layered convection does not take into account the energetic contribution of a possible phase separation. In that case, however, the energy release due to the phase separation does not need to be important, as the main role of immiscibility would be to trigger the double diffusive instability. Indeed, as both layered convection and phase separation increase the planet's cooling time, there is always a possible trade off between the respective contributions of these two processes to the planet's cooling rate. Note, however, that layered convection yields a hotter interior than the adiabatic evolution, favoring miscibility, especially in the past (see Supplementary \fig{fig:StructureSat}). As a smaller immiscibility region, if any, is predicted for helium in Jupiter\itref{LHR09}$^,$\itref{MSC09}, this scenario would also explain the dichotomy revealed here between Jupiter and Saturn.


The cooling history of our Solar System giants, and of extrasolar giant planets in general, might thus be more complex than assumed by the conventional, simplistic homogeneous, adiabatic interior structure and evolution paradigm. 
The amazing diversity of the observed properties of extrasolar planets - in particular the anomalously large observed radii of hot jupiters\itref{MF11} that cannot be fully explained by the various proposed heating mechanisms\itref{LCA11}$^,$\itref{LCB10} - 
combined with the luminosity anomalies of our own Solar System giants, clearly suggest a 
significant revision of this paradigm and point to a broader, more complex picture of solar and extrasolar giant planet structure, composition and thermal evolution, with a direct impact 
on giant planet formation conditions. 

\clearpage

\section*{Methods}

\subsection*{Layered convection and thermal gradient}

To study the impact of layered double diffusive convection on the interior thermal structure, we have developed an analytical model of the heat transport in such a medium\itref{LC12}, which is briefly outlined below. 

The fluid is assumed to be composed of a large number of small convective layers of height $\hl$, separated by thinner diffusive interfaces. The transport in the convective layers is described by a parametrization similar to the standard mixing length theory of convection, for which the mixing length is chosen to be equal to the height of the layers ($\hl$). The temperature gradient ($\delt\equiv \pd{\ln T}{\ln P}$), hence the superadiabaticity in the gas, $\delt-\delad$, is given as a function of the flux to be transported, of the thermodynamic properties of the medium and of $\hl$.

 In the diffusive interfaces, the thermal gradient is equal to the gradient needed to transport the whole intrinsic flux by diffusive processes alone ($\deld$). 
The thermal size of these diffusive interfaces, $\deltat$, can be estimated by the fact that the convective timescale, given by the inverse of the \brunt frequency ($\tau_\mathrm{c}=\Nt^{-1}$) must be equal to the diffusive timescale in the interface ($\tau_\mathrm{d}=\deltat^2/\kapt$, where $\kapt$ is the thermal diffusivity of the medium). Thus, $\deltat=\sqrt{\kapt/\Nt}$, and the only remaining free parameter is the height of the convective layers, or equivalently the mixing length parameter, $\alpha=\hl/\hp$, where $\hp$ is the pressure scale height in the fluid.

Finally, the mean temperature gradient $\delmean$ is a linear combination of the two aforementioned temperature gradients weighted by the relative size of the convective and diffusive regions, and depends on the size of the convective/diffusive cells\itref{LC12} (i.e. on $\alpha$).

\subsection*{Planetary model sequences and time integration}

Thermal evolution tracks of a planet of a given mass $\Mp$, composition (symbolically denoted by an array, $\tilde{X}$, encompassing all the information needed to know the abundances of all the materials at all depths, including the core), angular momentum ($\angmom$), and thermal structure (parametrized here by the mixing length parameter $\alpha$) yield relations of the type
\balign{\label{evol_equ}
&\Lint \equiv 4\pi\rp^2 \ssb \tint^4=4\pi\rp^2 \ssb \left(\te^4-\teq^4\right)=\Lint(\Mp,\tilde{X},\alpha,\angmom,t), \nonumber \\
&\etot\equiv \eth+\egrav+\erot=\etot(\Mp,\tilde{X},\alpha,\angmom,t),\nonumber \\
&\rp=\rp(\Mp,\tilde{X},\alpha,\angmom,t)  ,
}
where $\tint$ and $\te$ are the intrinsic and total effective temperatures, $\ssb$ the Stefan-Boltzmann constant and $\rp$ the radius of the planet.
The mixing length parameter is assumed constant as convective/diffusive layers are assumed to have reached a state of equilibrium\itref{LC12}$^,$\itref{Rad05}.  Without any further guidance from tridimensional hydrodynamical simulations covering a large enough spatial and temporal domain, it is sensible to start with the most simple assumption.

The equilibrium temperature ($\teq$) is the temperature that the planet would have if its intrinsic luminosity were zero and represents the contribution to the solar flux through
\balign{
\ssb \teq^4\equiv(1-A)\Lsun/(16\pi a^2),
}
where $\Lsun$ is the solar luminosity, $a$ the orbital distance of the planet, and $A$ its Bond albedo. 
Throughout this study, the values used for the equilibrium temperature are $\teq=109\,$K and $91\,$K for Jupiter and Saturn respectively. This corresponds to an absorbed luminosity of 5.14$\times 10^{16}\,$W (1.11$\times 10^{16}\,$W) and a Bond albedo of 0.37 (0.32) for Jupiter (Saturn). Within the observational uncertainties, those values are consistent with observations.

The relations described by (\ref{evol_equ}) can be derived with a grid of atmosphere models. These grids of atmospheric boundary conditions provide us with the temperature at the 10 bar level (or any reference level deeper than the photosphere), $\ttb(\tint,g)$, as a function of the intrinsic effective luminosity ($\tint$), and of the gravity, $g\equiv G\Mp/\rp^2$. The details of the functional form used to represent the atmospheric boundary conditions are detailed in the Supplementary Information.

Then, for any arbitrary value of the luminosity, $\ttb$ is known, and the $T-P$ profile is integrated inward by integrating the standard structure equations for a rotating body (Supplementary Information). Models are computed on a grid of luminosities. For each luminosity, the gravitational, thermal ($\uint$ being the specific internal energy) and rotational energy, 
\balign{
\egrav\equiv - \int{\frac{Gm}{\bar{r}}\d m}, \, \ \, 
\eth \equiv  \int{\uint}\, \d m, \, \ \,
\erot \equiv\frac{1}{2} \frac{\angmom^2}{\anginertia},
}
hence the total energy, are computed (see Supplementary \fig{fig:LtotEtotSat}). Rotational energy is computed assuming negligible angular momentum loss, and is found to represents only a small fraction of the total energy ($\approx 2\%$). As the ANEOS equation of state provides us with $\uint$ (the specific internal energy) for the volatiles, the thermal contribution of the core is included in our calculations.

Finally, we are left with a tabulated form of functions of the type $\etot=\etot(\Mp,\tilde{X},\alpha,\angmom,\Lint)$, and time dependence can be retrieved by choosing an initial value for the luminosity or initial energy content and by integrating 
\balign{-\Lint=\pdc{ \etot}{t}{\Mp,\tilde{X},\alpha,\angmom}.}

\clearpage

{\bf References and Notes}

\begin{enumerate}

\item \label{PGM77} Pollack, J. \textit{et al.} A calculation of Saturn's gravitational contraction history. {\it Icarus}. \textbf{30}, 111-128 (1977).

\item \label{FIN11} Fortney, J. J., Ikoma, M. Nettelmann, N., Guillot, T. \& Marley, M. S. Self-consistent Model Atmospheres and the Cooling of the Solar System's Giant Planets. \textit{Astrophys. J.} \textbf{729}, 32 (2011). 

\item \label{Sal73} Salpeter, E. E. On Convection and Gravitational Layering in Jupiter and in Stars of Low Mass. \textit{Astrophys. J.} \textbf{181}, L83-L86 (1973).

\item \label{SS77} Stevenson, D. J. \& Salpeter, E. E. The dynamics and helium distribution in hydrogen-helium fluid planets. \textit{ Astrophysical Journal Supplement Series}. \textbf{35}, 239-261 (1977b).


\item \label{HDW85} Hubbard, W. B. \& DeWitt, H. E. Statistical mechanics of light elements at high pressure. VII - A perturbative free energy for arbitrary mixtures of H and He. \textit{Astrophys. J.} \textbf{290}, 388-393 (1985)

\item \label{PHB95} Pfaffenzeller, O., Hohl, D. \& Ballone, P. Miscibility of Hydrogen and Helium under Astrophysical Conditions. \textit{Phys. Rev. Lett.} \textbf{74}, 2599-2602 (1995).

\item \label{LHR09} Lorenzen, W., Holst, B. \& Redmer, R. Demixing of Hydrogen and Helium at Megabar Pressures. \textit{Phys. Rev. Lett.} \textbf{102}, 115701 (2009).

\item \label{MSC09} Morales, M. A. \textit{et al.} Phase separation in hydrogen-helium mixtures at Mbar pressures. \textit{PNAS.} \textbf{106}, 1324 (2009).

\item \label{FH03} Fortney, J. J. \& Hubbard, W. B. Phase separation in giant planets: inhomogeneous evolution of Saturn. {\it Icarus}. \textbf{35} (2003).

\item \label{Ste85} Stevenson, D. J. Cosmochemistry and structure of the giant planets and their satellites. {\it Icarus}. \textbf{62}, 4-15 (1985).

\item \label{PHS91} Podolak, M., Hubbard, W. B. \& Stevenson, D. J. Model of Uranus' interior and magnetic field. In {\it Uranus}, University of Arizona Press, 29-61, (1991).

\item \label{CB07} Chabrier, G. \& Baraffe, I. Heat Transport in Giant (Exo)planets: A New Perspective. \textit{Astrophys. J. Lett.} \textbf{667}, L81-L84 (2007). 

\item \label{LC12} Leconte, J. \& Chabrier, G. A new vision of giant planet interiors: Impact of double diffusive convection. \textit{Astron. \& Astroph.} \textbf{540}, A20 (2012). 

\item \label{Ste60} Stern, M. E. The salt-fountain and thermohaline convection. {\it Tellus}. \textbf{12}, 172-+ (1960).

\item \label{Rad05} Radko, T. What determines the thickness of layers in a thermohaline staircase? \textit{J. Fluid Mech.} \textbf{523}, 79-98 (2005). 

\item \label{RGT11} Rosenblum, E. P., Garaud, P., Traxler, A. \& Stellmach, S. Turbulent Mixing and Layer Formation in Double-diffusive Convection: Three-dimensional Numerical Simulations and Theory. \textit{Astrophys. J.} \textbf{731},66-+ (2011). 

\item \label{MGS12} Mirouh, G. M., Garaud, P., Stellmach, S., Traxler, A. L. \& Wood, T. S. A New Model for Mixing by Double-diffusive Convection (Semi-convection). I. The Conditions for Layer Formation. \textit{Astrophys. J.} \textbf{750}, 61 (2012). 

\item \label{HK94} Hansen, C. J. \& Kawaler, S. D. \textit{Stellar Interiors. Physical Principles, Structure, and Evolution.} Springer-Verlag Berlin Heidelberg New York (1994). 

\item \label{PHB96} Pollack, J. B. \textit{et al.} Formation of the Giant Planets by Concurrent Accretion of Solids and Gas. \textit{Icarus}. \textbf{124}, 62 (1996).

\item \label{MFH07} Marley, M. S., Fortney, J. J., Hubickyj, O., Bodenheimer, P. \& Lissauer, J. J. On the Luminosity of Young Jupiters. \textit{Astrophys. J.} \textbf{655}, 541-549 (2007). 

\item \label{GSS} Guillot, T., Saumon, D. \& Stevenson, D. J. The Interior of Jupiter. In \textit{Jupiter. The Planet, Satellites and Magnetosphere.} 

\item \label{WM12} Wilson, H. F. \& Militzer, B. Solubility of Water Ice in Metallic Hydrogen: Consequences for Core Erosion in Gas Giant Planets. \textit{Astrophys. J.} \textbf{745}, 54 (2012).

\item \label{WM12b} Wilson, H. F. \& Militzer, B. Rocky Core Solubility in Jupiter and Giant Exoplanets. \textit{Phys. Rev. Lett.} \textbf{108}, 1101 (2012)

\item \label{Ste82} Stevenson, D. J. Formation of the giant planets. {\it Pla. Space Sci.}, \textbf{30}, 755-764 (1982).

\item \label{SG04} Saumon, D. \& Guillot, T. Shock Compression of Deuterium and the Interiors of Jupiter and Saturn. \textit{Astrophys. J.} \textbf{609}, 1170-1180 (2004)

\item \label{SS77a} Stevenson, D. J. \& Salpeter, E. E. The phase diagram and transport properties for hydrogen-helium fluid planets. \textit{Astrophys. J. Suppl. Ser.} \textbf{35}, 221-237 (1977a).

\item \label{MF11} Miller, N. \& Fortney, J. J. The Heavy-element Masses of Extrasolar Giant Planets, Revealed. \textit{Astrophys. J. Lett.} \textbf{736}, L29 (2011)

\item \label{LCA11} Laughlin, G., Crismani \& M., Adams, F. C. On the Anomalous Radii of the Transiting Extrasolar Planets. \textit{Astrophys. J. Lett.} \textbf{729}, L7+ (2011)

\item \label{LCB10} Leconte, J., Chabrier, G., Baraffe \& I., Levrard, B. Is tidal heating sufficient to explain bloated exoplanets? Consistent calculations accounting for finite initial eccentricity. \textit{Astron. \& Astroph.} \textbf{516}, A64+ (2010).


\end{enumerate}

Correspondence and requests for materials should be addressed to J.L.

\subsubsection*{Acknowledgments}
J.L. thanks J. Fortney for making his atmospheric grids available to us in electronic format. The research leading to these results has received funding from the European Research Council under the European Community's Seventh Framework Programme (FP7/2007-2013 Grant Agreement no. 247060)

\subsubsection*{Author contributions}

J.L. carried out analytical calculations, developed the model and performed the numerical simulations.
G.C. suggested the idea and carried out analytical calculations.
J.L and G.C. wrote the manuscript.

\clearpage


\section*{Supplementary Information}

\section{Layered convection or homogeneous double diffusive convection?}

The transport efficiency derived from numerical simulations\itref{RGT11}$^,$\itref{MGS12} for a homogeneous double diffusive medium (where transport is ensured by small scale turbulence) is equivalent, in our analytical model, to a medium in a state of layered convection with $\alpha\approx 10^{-8}-10^{-9}$ (see Figure 3 of Leconte and Chabrier\itref{LC12}).

The fact that measured gravitational moments constrain $\alpha$ to be larger than $10^{-6}$ (\itref{LC12}) thus seems to favor the occurrence of layered convection over homogeneous double diffusive convection in Solar System giant planets. In addition, the evolutionary constraint on the luminosity discussed in the present study, which constrain $\alpha$ to be larger larger than $10^{-4}$ (see \fig{fig:TeffSaturn_Msc}), also seems to justify this hypothesis. 

\section{Atmospheric model grids}

The relations described by (\ref{evol_equ}) can be derived with a grid of atmosphere models. These grids of atmospheric boundary conditions provide us with the temperature at the 10 bar level (or any reference level deeper than the photosphere), $\ttb(\tint,g)$, as a function of the intrinsic flux exiting the planet (parametrized by the intrinsic effective luminosity, $\tint$), and of the gravity, $g\equiv G\Mp/\rp^2$.

We use the recent atmospheric grids derived specifically for Jupiter and Saturn by Fortney et al.\itref{FIN11}. However, as a planet with layered convection can be quite hot (high internal temperature) but with a low luminosity (low effective temperature), our evolutionary tracks can start in the low $g$, low $\tint$ part of the parameter space that is usually not probed by adiabatic evolution models. The grids mentioned above thus do not cover this range and extrapolation in $g$ had to be used. The rather low dependence of $\ttb(\tint,g)$ on gravity justifies this procedure.

In order to retrieve the behavior of $\ttb(\tint,g)$ in both the high and low $\tint$ regime (arbitrarily separated at $\tmid$), the numerical grids where fitted separately in the two regimes by functions of the form
\balign{\ttb(\tint,g)=C+K g^{\beta}\left(\tint -T_0\right)^{\gamma}. \label{fitatm}}
A smooth function is recovered on the whole temperature range by interpolating linearly between the two functions in the range $\left[\tmid-\Delta T,\tmid+\Delta T\right]$. The 10 bar temperatures derived by this procedure do not differ by more than 2-3\% with respect to the tabulated values. Values used for the fitting parameters are given in \tab{tab:fitparam}.

\begin{table}[htb]
\begin{center}
\caption{Parameters used in \eq{fitatm} to fit the atmospheric grids of ref.\,\ref{FIN11}.}
\label{tab:fitparam}
\small
\begin{tabular}{cc|c|c|c} \hline\hline 
	&\multicolumn{2}{c}{\bf Jupiter} &\multicolumn{2}{c}{\bf Saturn}  \\ \hline
$\tmid$ (K)&\multicolumn{2}{c}{224}&\multicolumn{2}{c}{225}\\
$\Delta T$ (K)&\multicolumn{2}{c}{25}&\multicolumn{2}{c}{25}\\
&$T<\tmid$&$T>\tmid$&$T<\tmid$&$T>\tmid$\\
$C$ & 78.4 &  -283.  &  62.7  & -1680.  \\
$K$&  0.00550 &  198.  &  0.00202  &  831.  \\
$T_0$ &  -122.  &  143.  &  -97.1  &  141.  \\
$\beta$&  -0.182  &  -0.114  &  -0.180  &  -0.0721  \\
$\gamma$&  2.09  &  0.454  &  2.31  &  0.293  \\
\hline\hline
\end{tabular}
\normalsize
\end{center}
\end{table}

\section{Planetary structure integration}

When $\ttb$ is known, the $T-P$ profile is integrated inward by integrating the standard structure equations for a rotating body
\balign{
\pd{\,P}{\,m}&=-\frac{1}{4\pi}\frac{ G m}{ \bar{r}^4}+\frac{ \op^2}{6\pi \bar{r}}+\varphi_\omega(\bar{r}), \\
\pd{\,\bar{r}}{\,m}&=\frac{1}{4\pi \bar{r}^2\rho}, \\
\pd{\,T}{\,m}&= \frac{T}{P}\pd{\,P}{\,m}\delt, \label{chess:dtdm} 
}
where $m$ is the mass enclosed in the equipotential of mean radius $\bar{r}$, $\op$ the rotation rate of the planet, $\varphi_\omega(\bar{r})$ a second-order correction due to the centrifugal potential. As discussed above, $\delt$ is either equal to $\delad$ in the adiabatic case, or to $\delmean$ which depends on $\alpha$ in a layered convection zone.
The angular velocity is given by the fact that $\angmom\equiv \anginertia \op$, $\anginertia$ being the angular moment of inertia computed for each model, is kept constant and equal to the present value.

A closure equation is provided by the equation of state (EOS) of the mixture along the planet's interior profile. Such an EOS is generally given by the so-called ideal volume law for the mixture:
\balign{
\frac{1}{\rho}=\frac{X}{\rho_X}+\frac{Y}{\rho_Y}+\frac{Z}{\rho_Z},
}
where $X$, $Y$ and $Z$ denote the mass fractions of H, He and heavy elements, respectively.

For each value of $\alpha$, the core mass and the mass fraction of heavy elements as a function of depth are considered constant in time and equal to the distribution that best matches the gravitational moments and observational constraints\itref{LC12}.
For adiabatic, homogeneous models, the core masses and uniform metal enrichment in the envelope are equal to $\sim 4\,\mearth$ and $Z=0.11$ in Jupiter and $25.5\,\mearth$ and $Z=0.05$ in Saturn. For models of Saturn with layered convection, the metal enrichments used for a given mixing length parameter $\alpha$ are shown in \itref{LC12} and the core mass used is $\sim 20.5\,\mearth$. This difference in the core mass between the homogeneous model and the layered ones is responsible for the small energy difference remaining at low luminosity in Supplementary \fig{fig:LtotEtotSat}. 
Details about the procedure and the equation of states used can be found in ref.\,\ref{LC12} and reference therein.

Supplementary \fig{fig:TeffJupSat} shows adiabatic, homogeneous evolution tracks for Jupiter and Saturn computed with this method. As expected, in the homogeneous case, our calculations are in good agreement with previous results\itref{FIN11}: models cool to the observed effective temperature in $\sim2.9$\,Gyr for Saturn, and $\sim5.2$\,Gyr for Jupiter. As we use the same atmospheric model grids, the slight differences arising in the cooling times are probably due to the fact that our adiabatic models do not have exactly the same core mass as the most recent calculations\itref{FIN11}, and that we take into account the thermal contribution of this core.

\clearpage

\begin{figure}[htbp] 
 \centering
 \resizebox{.6\hsize}{!}{\includegraphics{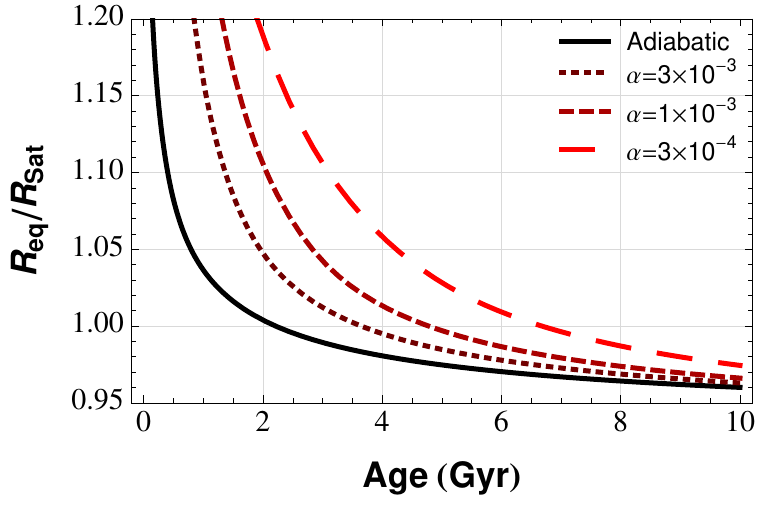}}
 \caption{
Temporal evolution of the equatorial radius (in units of the measured present equatorial radius of Saturn, 60\,268\,km) of an adiabatic Saturn model (solid black curve), and of three models with layered convection ($\alpha= 3\times10^{-3}$, $10^{-3}$ and $3\times10^{-4}$ from dark to light red). The effect of the deformation due to the fast rotation is taken into account. The time needed for the adiabatic model fulfilling the constraints on the measured gravitational moments to cool down and contract to the present radius is too short.}
 \label{fig:ReqSat}
\end{figure}
\begin{figure}[htbp] 
 \centering
 \resizebox{.6\hsize}{!}{\includegraphics{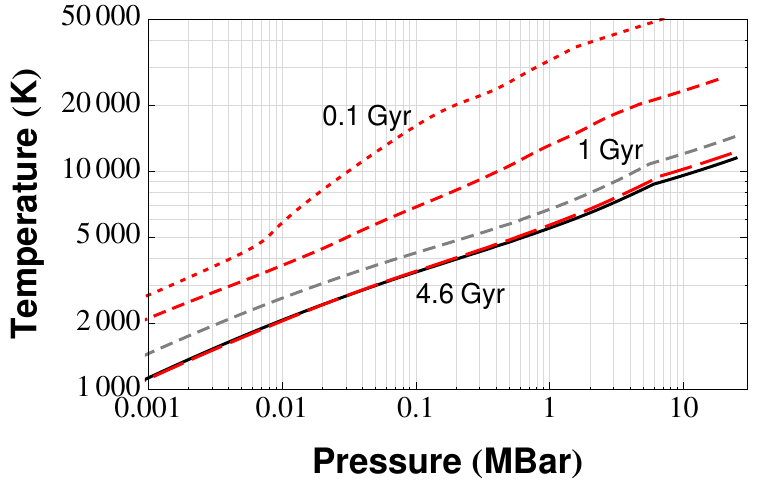}}
 \caption{
Internal temperature-pressure profile of our baseline model with layered convection (red curves; $M_\mathrm{LC}\approx0.8\,\Mp$; $\alpha=2\times10^{-3}$) at three different ages (dotted: 0.1\,Gyr; dashed: 1\,Gyr; long dashed: 4.6\,Gyr). The solid black curve represents the temperature profile inferred for the standard adiabatic model. For comparison, the 1\,Gyr profile of an adiabatic evolution is shown by the gray dashed curve.} 
 \label{fig:StructureSat}
\end{figure}
\begin{figure}[htbp] 
 \centering
 \resizebox{.6\hsize}{!}{\includegraphics{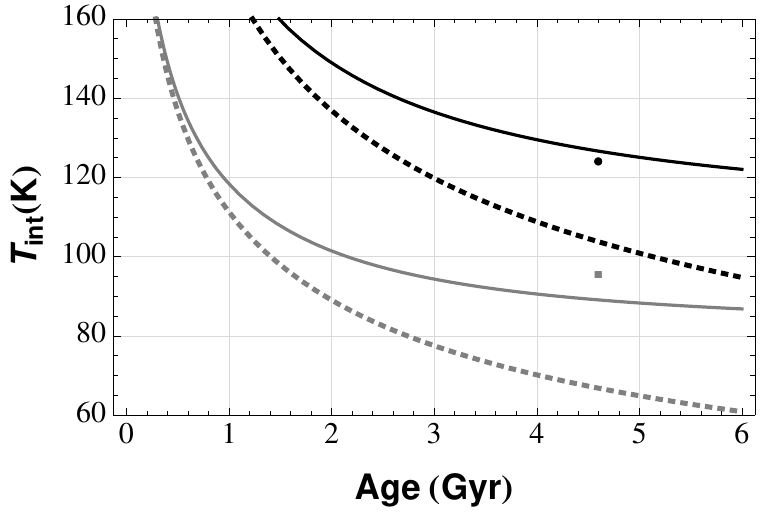}}
 \caption{
Effective ($\te$; solid curves) and intrinsic ($\tint$; dashed curves) temperature as a function of the planetary age for our adiabatic reference models of Jupiter (black) and Saturn (gray). Observed effective temperatures are also shown as dots. As seen, our calculations  in the conventional homogeneous adiabatic case are in good agreement with previous results\itref{FIN11}: models cool to the observed effective temperature in $\sim2.9$\,Gyr for Saturn, and $\sim5.2$\,Gyr for Jupiter.} 
 \label{fig:TeffJupSat}
\end{figure}
\begin{figure}[htbp] 
 \centering
 \resizebox{.6\hsize}{!}{\includegraphics{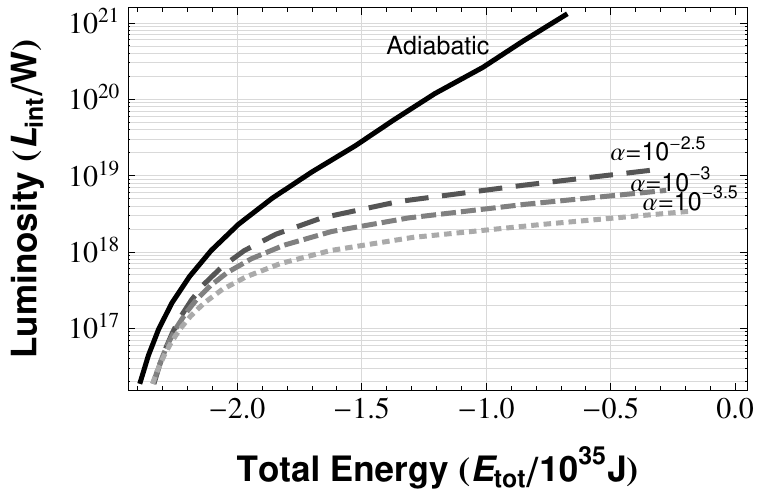}}
 \caption{
Intrinsic luminosity ($\Lint\equiv 4\pi \rp^2 \,\fint$) as a function of the total internal energy ($\etot\equiv\egrav + \eth + \erot $) for various sequences of Saturn models. The solid curve is the reference adiabatic model. The long dashed, dashed and dotted curves are the $\alpha= 10^{-2.5}$, $10^{-3}$ and $10^{-3.5}$ sequences, respectively. As expected, for a given internal energy content, the intrinsic luminosity of models assuming layered convection decreases when $\alpha$ decreases and is significantly reduced compared with the adiabatic  case. This increases the Kelvin Helmholtz timescale, $\tKH\equiv G\Mp^2/(\rp\Lint)$, i.e. the time needed for the object to loose the memory of its initial energy state.}
 \label{fig:LtotEtotSat}
\end{figure}

\end{document}